\documentclass[conference, twocolumn]{IEEEtran}

\usepackage{cite}
\usepackage{amsmath,amssymb,amsfonts}
\usepackage{gensymb}
\usepackage{float}
\usepackage{stfloats}

\usepackage{array}
\usepackage{algorithmic}
\usepackage{graphicx}
\usepackage{textcomp}
\usepackage{xcolor}
\usepackage[hidelinks]{hyperref}
\usepackage{cancel}

\usepackage{amsthm}
\theoremstyle{definition}

\ifCLASSOPTIONcompsoc
    \usepackage[caption=false, font=normalsize, labelfont=sf, textfont=sf]{subfig}
\else
\usepackage[caption=false, font=footnotesize]{subfig}
\fi

\DeclareMathOperator{\sinc}{sinc}

\begin{document}

\title{Analysis of the Range Ambiguity Function of Narrowband Near-field MIMO Sensing}
 
\author{
    \IEEEauthorblockN{Marcin Wachowiak\IEEEauthorrefmark{1}\IEEEauthorrefmark{2}, 
    André Bourdoux\IEEEauthorrefmark{1}, 
    Sofie Pollin\IEEEauthorrefmark{2}\IEEEauthorrefmark{1}}
    \IEEEauthorblockN{
        \IEEEauthorrefmark{1} imec, Kapeldreef 75, 
        3001 Leuven, Belgium \\
        \IEEEauthorrefmark{2} Department of Electrical Engineering, 
        KU Leuven, Belgium \\
        Email: marcin.wachowiak@imec.be 
        }   
}

\maketitle

\begin{abstract}
This paper compares the sensing performance of a narrowband near-field system across several practical antenna array geometries and SIMO/MISO and MIMO configurations. 
For identical transmit and receive apertures, MIMO processing is equivalent to squaring the near-field array factor, resulting in reduced beamdepth and improved sidelobe level.
Numerical results demonstrate that MIMO processing improves the near-field resolution and maximum sensing range by approximately a factor of 1.4 compared to a single-aperture system. 
Using a quadratic approximation of the mainlobe of the array factor, an analytical improvement factor of $\sqrt{2}$ is derived, validating the numerical results. 
Finally, MIMO is shown to improve the poor sidelobe performance observed in the near-field by a factor of two, due to squaring of the array factor.
\end{abstract}

\begin{IEEEkeywords}
Radiative near-field, finite-depth beamforming, near-field beamforming, near-field sensing, MIMO
\end{IEEEkeywords}

\section{Introduction}

\subsection{Problem Statement}
The radiative near-field (NF) systems are regarded as a key enabler of the next generation of wireless systems \cite{6g_large_arr_tutorial}. Their ability to create beamspots in the distance domain enables user multiplexing both in range and angle \cite{nf_beamfocusing_comms, large_arr_beamdepth}. This unique property has also attracted the attention of the sensing community as beamfocusing in the range domain can be utilized to localize the targets with limited bandwidth \cite{nf_sensing_testbed}.
Despite these advantages, near-field sensing performance is subject to significant shortcomings, such as limited resolution, short maximum near-field range and low peak-to-sidelobe level (PSL).
The recent research regarding near-field sensing has primarily focused on single-array geometry and single-aperture systems \cite{primer_on_nf_bf, large_arr_beamdepth, uca_near_field, nf_sensing_testbed}.
To enhance the near-field sensing, it is essential to compare the performance across different antenna array geometries. Furthermore, the gains from MIMO processing in the near-field must be identified and quantified for each configuration.

\subsection{Relevant works}
The benefits of MIMO processing for radar and a far-field scenario were investigated in \cite{mimo_radar}.
In \cite{primer_on_nf_bf}, the near-field beam focusing was derived for a uniform square array, based on electromagnetic principles. 
Generalizations of the NF array factor (AF) of a uniform rectangular array (URA) and a uniform planar circular array (UPCA) are provided in \cite{large_arr_beamdepth}. The work studied the beamfocusing property from a communication-centric perspective.
The near-field communications performance and AF of a uniform circular array (UCA) were discussed in \cite{uca_near_field}.
In \cite{nf_sensing_testbed}, the near-field sensing resolution was demonstrated using a testbed equipped with a uniform linear antenna array (ULA).
The sensing performance of a near-field system for point and extended targets was investigated in \cite{nf_af_pt_et_miso_simo}. The work discusses the NF performance of the SIMO/MISO and MIMO systems regarding the Cramér–Rao lower bound, but does not directly discuss the gains from MIMO for point targets.

\begin{figure}[t]
    \centering
    \includegraphics[width=\linewidth]{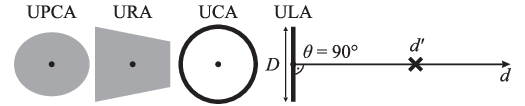}
    \caption{System diagram.}
    \label{fig:sys_diagram}
\end{figure}

\subsection{Contributions}
This work evaluates and compares the narrowband NF sensing performance across several array geometries: ULA, UCA, URA and UPCA. To the best of our knowledge, this is the first work to compare the NF AF and beamdepth across different antenna array geometries. The NF sensing performance is studied for two cases: SIMO/MISO - single aperture and MIMO - two identical collocated apertures. The gains from employing MIMO are assessed in terms of resolution, maximum NF range
and PSL of the ambiguity function. Based on a quadratic approximation, the analytical improvement factor is derived, which matches well with the numerical simulations.

\section{Signal model}

\subsection{Ambiguity function of near-field sensing system}

Consider a collocated narrowband antenna array with $M$ transmitters and $N$ receivers. The location of the of the $m$th transmitter is given by $\mathbf{p}_m^{\mathrm{tx}} = \left[ x_m^{\mathrm{tx}}, y_m^{\mathrm{tx}}, z_m^{\mathrm{tx}} \right]^T$ and $n$th receiver by $\mathbf{p}_n^{\mathrm{rx}} = \left[ x_n^{\mathrm{rx}}, y_n^{\mathrm{rx}}, z_n^{\mathrm{rx}}\right]^T$. The antenna elements are assumed to be isotropic and mutual coupling is considered negligible.
The distance between each antenna element and some point in space $\mathbf{p} =\left[x, y, z\right]^T$ is 
$d_m^{\mathrm{tx}}(\mathbf{p}) = \left|\left| \mathbf{p} - \mathbf{p}_m \right|\right|_2 $ for transmitters and 
$d_n^{\mathrm{rx}}(\mathbf{p}) = \left|\left| \mathbf{p} - \mathbf{p}_n \right|\right|_2 $ for receivers.
The bistatic distance is denoted as $d_{m,n}(\mathbf{p}) = d_m^{\mathrm{tx}}(\mathbf{p}) + d_n^{\mathrm{rx}}(\mathbf{p})$.
The target is located within the radiative near-field of the array where the path loss variations across the aperture can be considered negligible \cite{primer_on_nf_bf} and the propagation is modeled as a line of sight. Under the assumption that the complex target response is invariant across all transmit-receive paths and the channel simplifies to phase shift
\begin{align}
    h_{m,n}(\mathbf{p}) &= e^{-j 2\pi \frac{f}{c} d_{m,n}(\mathbf{p}) }.
\end{align}
Given the channel model and array configuration, the matched filter response (MF) or ambiguity function for a single frequency and a target at position $\mathbf{p}'$ is
\begin{align}
    \mathcal{A}(\mathbf{p}', \mathbf{p}) &= \frac{1}{\sqrt{M N}} \sum_{m \in M} \sum_{n \in N}  h_{m,n}(\mathbf{p}') h_{m,n}^*(\mathbf{p}).
\end{align}
Note that the formulated ambiguity function does not consider bandwidth and is computed for a single frequency. This is a valid assumption for narrowband systems in which the resolution due to bandwidth is much smaller than the one provided by the NF beamdepth.
The ambiguity function can be separated into the product of two sums as follows
\begin{align}
    \label{eq:nf_af_mimo_full_form}
    \mathcal{A}(\mathbf{p}', \mathbf{p}) 
    &= \frac{1}{\sqrt{M}} \sum_{m \in M} e^{-j 2\pi \frac{f}{c} \Delta d_{m}^{\mathrm{tx}}(\mathbf{p}', \mathbf{p}) } \nonumber \\
    & \quad \times \frac{1}{\sqrt{N}} \sum_{n \in N}  e^{-j 2\pi \frac{f}{c} \Delta d_{n}^{\mathrm{rx}}(\mathbf{p}', \mathbf{p}) }.
\end{align}
where $\Delta d_m^{\mathrm{tx}}(\mathbf{p}', \mathbf{p}) = d_m^{\mathrm{tx}}(\mathbf{p}') - d_m^{\mathrm{tx}}(\mathbf{p})$ and $\Delta d_n^{\mathrm{rx}}(\mathbf{p}', \mathbf{p}) = d_n^{\mathrm{rx}}(\mathbf{p}') - d_n^{\mathrm{rx}}(\mathbf{p})$ is the distance difference between $\mathbf{p}'$ and $\mathbf{p}$ for $m$th and $n$th antenna.
The per antenna distance difference can be expressed as a distance difference with regard to the reference antenna element ($m=0$ and $n=0$) and a distance difference dependent on the antenna index
\begin{align}
    \label{eq:distance_rel_arr_center}
    \Delta d_{m}^{\mathrm{tx}}(\mathbf{p}', \mathbf{p}) &= \Delta d_{m=0}^{\mathrm{tx}}(\mathbf{p}', \mathbf{p}) + \Delta\delta d_m^{\mathrm{tx}}(\mathbf{p}', \mathbf{p}) \\
    \Delta d_{n}^{\mathrm{rx}}(\mathbf{p}', \mathbf{p}) &= \Delta d_{n=0}^{\mathrm{rx}}(\mathbf{p}', \mathbf{p}) + \Delta\delta d_n^{\mathrm{rx}}(\mathbf{p}', \mathbf{p}).
\end{align}
By expanding \eqref{eq:nf_af_mimo_full_form} with \eqref{eq:distance_rel_arr_center}, the matched filter can be expressed in terms of array factors as 
\begin{align}
    &\mathcal{A}(\mathbf{p}', \mathbf{p}) =  
    e^{-j 2\pi \frac{c}{f} \Delta d_{m=0}^{\mathrm{tx}}(\mathbf{p}', \mathbf{p})} 
    \underbrace{\frac{1}{\sqrt{M}} \sum_{m \in M} e^{-j 2\pi \frac{f}{c} \Delta\delta d_m^{\mathrm{tx}}(\mathbf{p}', \mathbf{p}) }}
    _{\mathrm{AF_{TX}}(\mathbf{p}', \mathbf{p})} \nonumber \\
    & \qquad \times  
    e^{-j 2\pi \frac{c}{f} \Delta d_{n=0}^{\mathrm{rx}}(\mathbf{p}', \mathbf{p})} 
    \underbrace{\frac{1}{\sqrt{N}}
    \sum_{n \in N} e^{-j 2\pi \frac{f}{c} \Delta\delta d_n^{\mathrm{rx}}(\mathbf{p}', \mathbf{p}) }}_{\mathrm{AF_{RX}}(\mathbf{p}', \mathbf{p})}.
\end{align}
When computing the power of the MF, the phase offsets are removed, leading to
\begin{align}
    |\mathcal{A}(\mathbf{p}', \mathbf{p})|^2 &= |\mathrm{AF_{TX}}(\mathbf{p}', \mathbf{p}) ) |^2
    |\mathrm{AF_{RX}}(\mathbf{p}', \mathbf{p})|^2.
\end{align}
Similarly, as in the far-field \cite{mimo_radar}, the MIMO configuration improves the ambiguity function by introducing a multiplication by a second array factor, in contrast to the  SIMO/MISO case, where only a single array factor defines the ambiguity function.
However, in the near-field region, the AF depends both on angle and distance from the array, unlike in the far-field, where it is only a function of angle. The distance sensitivity results in an array gain that varies with distance and enables beamfocusing. The distance-dependent beampattern allows for the discrimination of targets and users located at the same angle but at different ranges, without bandwidth, which is impossible in the far field.

\subsection{Near-field 3 dB beamdepth}
Consider a target and points of interest located along the array broadside, the vectors $\mathbf{p}$ and $\mathbf{p}'$ are reduced to scalars $d$ and $d'$, which represent the distance from the geometric center of the array along the normal. Fig. \ref{fig:sys_diagram} illustrates the system setup.
The AF in the near-field can be calculated analytically for different array architectures. \cite{primer_on_nf_bf, large_arr_beamdepth, uca_near_field}.
Due to a second-order Taylor expansion (the Fresnel approximation) of distance, the argument of the AFs shares the same formula for different array geometries
\begin{align}
    \label{eq:nf_af_arg}
    x(d', d) &= a d_{\mathrm{FA}} d_{\mathrm{ver}}
\end{align}
where $a$ is an argument scaling coefficient dependent on the array geometry, $d_{\mathrm{FA}} = 2D^2 / \lambda$ is the Fraunhofer distance of the array, $D$ is the array aperture, $\lambda$ is the wavelength and $d_{\mathrm{ver}} = \left| \frac{1}{d'}- \frac{1}{d} \right| = \frac{|d - d'|}{d'd}$ is the absolute value of vergence difference \cite{geom_optics}.
For brevity in the following, the shorthand $x$ is used to represent $x(d,d')$.
Consider some near-field ambiguity function given by $|\mathcal{A}(x)|^2$. 
For a SISO/SIMO system, the ambiguity function will be defined by a single array factor of the transmit or receive aperture $|\mathcal{A}(x)|_{\mathrm{SIMO/MISO}}^2 = |\mathrm{AF}(x)|^2$.
For the MIMO scheme, the ambiguity function will be a product of two array factors. For simplicity, the transmit and receive arrays are considered to be identical and collocated, resulting in a monostatic system $|\mathcal{A}(x)|_{\mathrm{MIMO}}^2 = |\mathrm{AF}(x)|^4$. 

The range resolution is evaluated in terms of the radial half-power beamwidth of the ambiguity function, also referred to as the beamdepth (BD). Note that in the considered narrowband system, the primary source of resolution is the distance-dependent beampattern. The resolution due to bandwidth is considered to be negligibly small compared to the resolution due to beamfocusing. The 3 dB beamdepth is calculated by computing the argument for which the ambiguity function reaches half of the maximum value. Considering a normalized ambiguity function it can be written as $|\mathcal{A}(x_{\mathrm{3 dB}})|^2 = 0.5$.
The argument $x_{\mathrm{3dB}}$ is then used to calculate the distance $d_{\mathrm{3dB}}$ at which the $-3$ dB value is reached by solving \eqref{eq:nf_af_arg} resulting in
\begin{align}
    \label{eq:3dB_distance}
    d_{\mathrm{3 dB}} = \frac{d_{\mathrm{FA}} d'}{ d_{\mathrm{FA}} \pm \alpha d'}.
\end{align}
Formula \eqref{eq:3dB_distance} gives two distances for which the $-3$ dB value is achieved for the target at $d'$. Where $\alpha = x_{\mathrm{3 dB}} / a$ represents the value of the argument $d_{\mathrm{FA}} d_{\mathrm{ver}}$ for which the normalized AF reaches half of the maximum value.
The parameter $a$ depends only on the array geometry, while $x_{\mathrm{3 dB}}$ depends on both the array architecture and configuration, SIMO/MISO or MIMO.

As $d_{\mathrm{3 dB}}$ should always be positive, the following constraint arises $d' < \frac{d_{\mathrm{FA}}}{ \alpha }$.
Given the $-3$ dB crossing points, 3 dB beamdepth (BD) is 
\begin{align}
    \mathrm{BD} 
    &= 
    \begin{cases}
        \frac{2 \alpha  d_{\mathrm{FA}} d'^2}{d_{\mathrm{FA}}^2 - \alpha^2 d'^2}, &d' < \frac{d_{\mathrm{FA}}}{\alpha},  \\
        \infty, & d' \geq \frac{d_{\mathrm{FA}}}{\alpha}.
    \end{cases}
\end{align}
For $d$ greater or equal to $\frac{d_{\mathrm{FA}}}{\alpha}$ the beamdepth extends from $d_{\mathrm{FA}} / \alpha$ to infinity. In the context of beamfocusing, the maximum near-field range is defined by a maximum distance for which the beamdepth remains finite, which is $d_{\mathrm{FA}} / \alpha$. Note that parameter $\alpha$ determines both the beamdepth and the maximum beamfocusing range, with smaller alpha corresponding to better performance.
In the following sections, parameter $\alpha$ is computed for different array architectures to assess the resolution and maximum near-field sensing range. 
The target is assumed to be located on the array broadside, corresponding to the best-case scenario, which offers the highest resolution due to the largest effective aperture. For angles other than the broadside, the effective aperture size is decreased and can be approximated as a projection of the array on a plane that is perpendicular to the axis of the target \cite{large_arr_beamdepth}. For example, for ULA, the effective aperture for a target at azimuth angle $\theta$ is $D_{\mathrm{eff}} = D \sin{(\theta)}$. For UCA, the aperture is constant for any $\theta$ due to circular symmetry of the array. For planar arrays, computation of the effective aperture becomes more cumbersome as it depends on the target elevation angle.

\section{Results}

\subsection{Uniform Linear Array (ULA)}
Consider a uniform linear array with element spacing lower than or equal to $\lambda / 2$. For points at the broadside of the array, the NF AF can be approximated as \cite{large_arr_beamdepth} 
\begin{align}
    &\left| \mathrm{AF}_{\mathrm{ULA}}(d', d) \right|^2 =  \\ 
    & \quad = M \frac{4}{ d_{\mathrm{FA}} d_{\mathrm{ver}} }
    \left( 
    C^2\left( \sqrt{\frac{ d_{\mathrm{FA}} d_{\mathrm{ver}}}{4}} \right) + S^2\left( \sqrt{\frac{ d_{\mathrm{FA}} d_{\mathrm{ver}}}{4}} \right) 
    \right), \nonumber
\end{align}  
where $C(u) = \int_{0}^{u} \cos{\left(\pi t^2 / 2 \right)} \, dt $ and $S(u) = \int_{0}^{u} \sin{\left(\pi t^2 / 2 \right)} \, dt$ are Fresnel integrals. For ULA the combined argument of the AF is $x = \frac{d_{\mathrm{FA}} d_{\mathrm{ver}}}{4}$ and hence $a = 1/4$.

\subsubsection{SIMO/MISO}
The normalized AF is 
\begin{align}
    |\mathrm{AF}(x)|^2 = \frac{C^2(\sqrt{x}) + S^2(\sqrt{x})}{x}.
\end{align}
The $-3$ dB value is achieved for argument $x_{\mathrm{3 dB}} = 1.738$, which results in $\alpha = 6.952$.

\subsubsection{MIMO}
The normalized AF is 
\begin{align}
    |\mathrm{AF} (x)|^4 = \left(\frac{C^2(\sqrt{x}) + S^2(\sqrt{x})}{x}\right)^2.
\end{align}
The $-3$ dB value is achieved for argument $x_{\mathrm{3 dB}} = 1.242$, which results in $\alpha = 4.968$.

\subsection{Uniform Circular Array (UCA)}
Consider a uniform circular array with antenna elements distributed on a circle of diameter $D$. Given that the spacing between the antennas is lower than or equal to $\lambda/2$, the NF AF can be approximated as \cite{uca_near_field}
\begin{align}
    \left| \mathrm{AF}_{\mathrm{UCA}}(d', d) \right|^2 &= M \left( J_0 \left( \frac{\pi d_{\mathrm{FA}} d_{\mathrm{ver}}}{16} \right) \right)^2,
\end{align}
where $J_0$ denotes the zero-order Bessel function of the first kind.
For UCA the combined argument of the AF is $x= \frac{\pi d_{\mathrm{FA}} d_{\mathrm{ver}}}{16}$ and hence $a = \pi / 16$. 

\subsubsection{SIMO/MISO}
The normalized AF is
\begin{align}
    |\mathrm{AF}(x)|^2 = \left(J_0(x) \right)^2.
\end{align}
The $-3$ dB value is achieved for argument $x_{\mathrm{3 dB}} = 1.126$, which results in $\alpha = 5.735$.

\subsubsection{MIMO}
The normalized AF is 
\begin{align}
    |\mathrm{AF}(x)|^4 = & \left(J_0(x) \right)^4.
\end{align}
The $-3$ dB value is achieved for argument $x_{\mathrm{3 dB}} = 0.815$, resulting in $\alpha = 4.151$.
    
\subsection{Uniform Rectangular Array (URA)}
Consider a uniform rectangular array with equal planar dimensions, resulting in a square array. In a square array, the total aperture is measured as the diagonal and equals $D$. For antenna spacing lower than or equal to $\lambda/2$ and points on the array normal, the NF AF can be approximated as \cite{large_arr_beamdepth}
\begin{align}
    &\left| \mathrm{AF}_{\mathrm{URA}} (d', d) \right|^2 = \\
    & \quad  M \left( \frac{8}{ d_{\mathrm{FA}} d_{\mathrm{ver}} } \right)^2
    \left( C^2\left( \sqrt{\frac{ d_{\mathrm{FA}} d_{\mathrm{ver}}}{8}} \right) 
    + S^2\left( \sqrt{\frac{ d_{\mathrm{FA}} d_{\mathrm{ver}}}{8}} \right) \right)^2. \nonumber  
\end{align}
For square URA the combined argument of the AF is $x = \frac{d_{\mathrm{FA}} d_{\mathrm{ver}}}{8}$ and hence $a = 1/8$.

\subsubsection{SIMO/MISO}
The normalized AF is 
\begin{align}
    |\mathrm{AF} (x)|^2 = & \left( \frac{C^2(\sqrt{x}) + S^2(\sqrt{x})}{x} \right)^2.
\end{align}
The $-3$ dB value is achieved for argument $x_{\mathrm{3 dB}} = 1.242$, which results in $\alpha = 9.936$.

\subsubsection{MIMO}
The normalized AF is 
\begin{align}
    |\mathrm{AF} (x)|^4  =  \left( \frac{C^2(\sqrt{x}) + S^2(\sqrt{x})}{x} \right)^4.
\end{align}
The $-3$ dB value is achieved for argument $x_{\mathrm{3 dB}} = 0.884$, which results in $\alpha = 7.072$.

\subsection{Uniform Planar Circular Array (UPCA)}
Consider an array implemented as a uniform planar circular array constructed from concentric circles. Given that the spacing between the antenna elements does not exceed $\lambda/2$, the NF AF can be approximated as \cite{large_arr_beamdepth}
\begin{align}
    \left| \mathrm{AF}_{\mathrm{UPCA}}(d', d) \right|^2 = M \sinc^2{\left( \frac{d_{\mathrm{FA}} d_{\mathrm{ver}}}{16} \right)},
\end{align}
where $\sinc{(x)} = \sin{(\pi x)} / (\pi x)$.
For UPCA the combined argument of the AF is $x = \frac{d_{\mathrm{FA}} d_{\mathrm{ver}}}{16}$ and hence $a = 1/16$.

\subsubsection{SIMO/MISO}
The normalized AF is
\begin{align}
    |\mathrm{AF} (x)|^2 = \sinc^2{(x)}.
\end{align}
The $-3$ dB value is achieved for argument $x_{\mathrm{3 dB}} = 0.443$, which results in $\alpha = 7.088$.
      
\subsubsection{MIMO}
The normalized AF is
\begin{align}
    |\mathrm{AF} (x)|^4 = \sinc^4{(x)}.
\end{align}
The $-3$ dB value is achieved for argument $x_{\mathrm{3 dB}} = 0.319$, which results in $\alpha = 5.104$. 

\begin{table}[b]
    \caption{Value of $\alpha$ parameter per array geometry and processing.}
    \label{tab:alpha_per_geom}
    \centering
    \def\arraystretch{1.2}
    \begin{tabular}{
      >{\centering\arraybackslash}m{0.20\linewidth}<{}
      |>{\centering\arraybackslash}m{0.18\linewidth}<{}
      |>{\centering\arraybackslash}m{0.13\linewidth}<{}
      |>{\centering\arraybackslash}m{0.27\linewidth}<{}
    }
         Array geometry & $\alpha_{\mathrm{SIMO/MISO}}$ & $\alpha_{\mathrm{MIMO}}$ & Ratio $\frac{\alpha_{\mathrm{SIMO/MISO}}}{\alpha_{\mathrm{MIMO}}}$ \\
         \hline
         ULA & 6.952 & 4.969 & 1.399\\
         \hline
         UCA & 5.737 & 4.148 & 1.383\\
         \hline
         URA & 9.937 & 7.068 & 1.406\\
         \hline
         UPCA & 7.087 & 5.103 & 1.389\\
         
    \end{tabular}
\end{table}

\begin{figure}[b!]
    \centering
    \includegraphics[width=\linewidth]{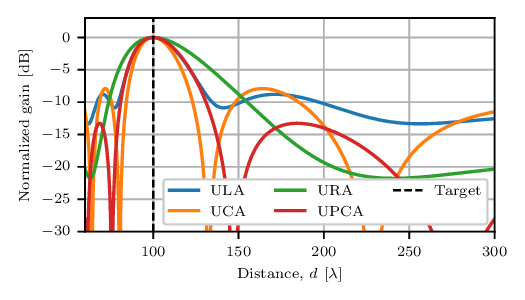}
    \caption{Array factor per geometry for SIMO/MISO, $ D=50\lambda$ and target at $d'= 100 \lambda$.}
    \label{fig:af_per_geom_simo_miso}
\end{figure}

\begin{figure}[b]
    \centering
    \includegraphics[width=\linewidth]{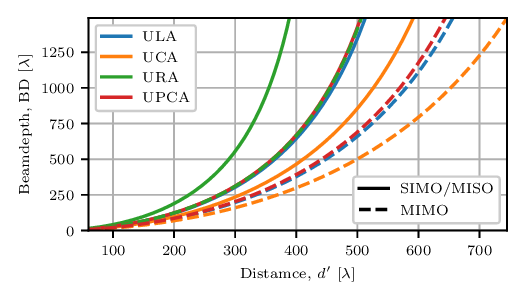}
    \caption{Beamdepth vs distance for different antenna array geometries and configurations}
    \label{fig:bd_vs_d_per_geom_simo_miso_vs_mimo}
\end{figure}

\begin{figure*}[t]
        \centering
        \subfloat[ULA \label{fig:af_ula}]{
            \includegraphics[width=0.475\linewidth]{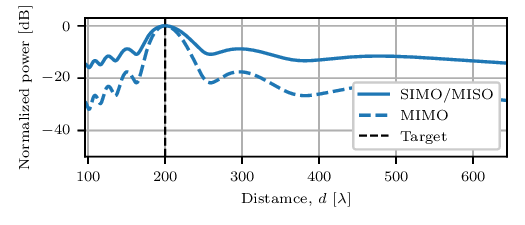}
        }
        \hfill
        \subfloat[UCA \label{fig:af_uca}]{
            \includegraphics[width=0.475\linewidth]{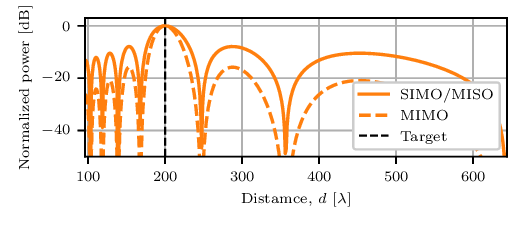}    
        }
        \\
        \subfloat[URA  \label{fig:af_ura}]{
            \includegraphics[width=0.475\linewidth]{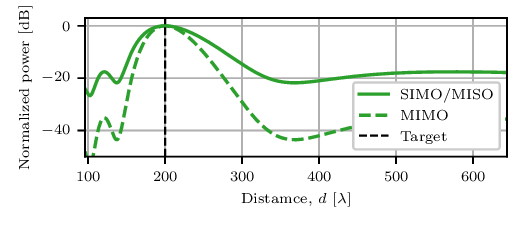}
        }
        \hfill
        \subfloat[UPCA \label{fig:af_upca}]{
            \includegraphics[width=0.475\linewidth]{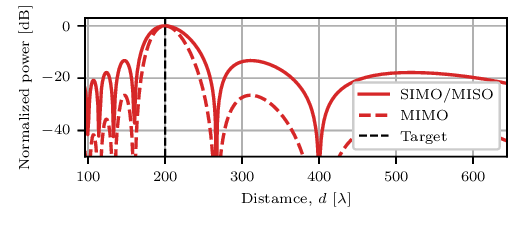}
        }        
        \caption{Ambiguity function for selected antenna arrays and processing for $D = 80 \lambda$ and target at $d'= 200\lambda$. } 
        \label{fig:af_per_geom}
\end{figure*}

Table \ref{tab:alpha_per_geom} presents the summarized values of the parameter $\alpha$ for each considered array geometry and processing. 
By comparing the $\alpha$ values across antenna geometries and processing a few observations can be made. For the same array aperture, UCA offers the best resolution, while requiring approximately $\pi$ times more antennas than the ULA. The URA provides the widest beamdepth due to its aperture being defined by the diagonal. This results in reduced effective aperture extent along the X and Y axes. 
Fig. \ref{fig:af_per_geom_simo_miso} illustrates the SIMO/MISO AF per geometry for the target at the same distance. The 1D arrays offer the best resolution with a low antenna count but exhibit poor sidelobe performance.
Fig. \ref{fig:bd_vs_d_per_geom_simo_miso_vs_mimo} shows the BD as a function of the distance for different configurations. The BD function rapidly reaches high values; however, employing MIMO improves the BD across the entire considered NF region.

The MIMO processing improves the parameter $\alpha$ approximately by the same factor of $1.4$ regardless of the array geometry. This corresponds to $1.4$-fold improvement in the NF sensing resolution and maximum NF range.
The consistent ratio $\frac{\alpha_{\mathrm{SIMO/MISO}}}{\alpha_{\mathrm{MIMO}}}$ observed across various array geometries can be attributed to quadratic decay of the AF within the mainlobe. Based on this, the mainlobe of the AF can be accurately approximated as a quadratic function \cite{quad_approx_mf}. For small arguments, the NF AF mainlobe is an even function, which for SIMO/MISO can be approximated as
\begin{align}
    \label{eq:af_quad_approx}
    |\mathrm{AF}(x)|^2 &\approx 1 - cx^2,
\end{align}
where $c$ is some scalar that controls the decay rate and depends on the array geometry.
Similarly, as in the previous subsections, the $-3$ dB value is obtained for the argument   
$x_{\mathrm{3 dB}}^{\mathrm{SIMO/MISO}} = \frac{\sqrt{2}}{2\sqrt{c}}$. Note that the approximation is valid only for the argument $x$ within the mainlobe.
In the MIMO case, the approximation from  \eqref{eq:af_quad_approx} is squared. To keep the approximation quadratic for both scenarios, higher order terms are neglected, resulting in 
\begin{align}  
    |\mathrm{AF}(x)|^4 &\approx 1 - 2cx^2.
\end{align}
The $-3$ dB value is obtained for argument $x_{\mathrm{3 dB}}^{\mathrm{MIMO}} = \frac{1}{2\sqrt{c}}$.
Consequently, the ratio between the SIMO/MISO and MIMO arguments is
\begin{align}
    \frac{x_{\mathrm{3 dB}}^{\mathrm{SIMO/MISO}}}{x_{\mathrm{3 dB}}^{\mathrm{MIMO}}} = \sqrt{2}.
\end{align}
According to the quadratic approximation, introducing the MIMO scheme should reduce the $x_{\mathrm{3 dB}}$ argument by a factor $\sqrt{2}$, which corresponds to $\frac{\alpha_{\mathrm{SIMO/MISO}}}{\alpha_{\mathrm{MIMO}}}$ equal to $\sqrt{2}$. Notice that the parameter $c$, which controls the rate of the decay of the AF mainlobe, does not affect the ratio of the arguments. This confirms the observation from Table \ref{tab:alpha_per_geom} that, although the rate of decay varies across array geometries, the improvement in the argument offered by MIMO is identical. The worst-case relative error introduced by the approximation is below $2.27\%$, matching well with the numerical results.
Fig. \ref{fig:af_per_geom} illustrates the ambiguity function for each considered array geometry and processing. As anticipated, MIMO processing improves the resolution by offering an array factor that decays at a faster rate. Note that the nulls of the ambiguity function remain the same regardless of the processing, as squaring any arbitrary real function preserves the locations of the zeros. 

Moreover, the MIMO processing significantly improves the peak-to-sidelobe level. Squaring of the AF corresponds to multiplication of the decay rate by 2, resulting in a twofold improvement of the PSL, as seen in Fig. \ref{fig:af_per_geom}. Table \ref{tab:psl_per_geom} presents the PSL per antenna geometry and processing. Interestingly, the arrays distributed on a plane (URA, UPCA) offer much better PSL than the one-dimensional arrays (ULA, UCA), although the latter provide better resolution.

\begin{table}[t]
    \caption{Peak-to-sidelobe level per array geometry and processing.}
    \label{tab:psl_per_geom}
    \centering
    \def\arraystretch{1.2}
    \begin{tabular}{
      >{\centering\arraybackslash}m{0.20\linewidth}<{}
      |>{\centering\arraybackslash}m{0.32\linewidth}<{}
      |>{\centering\arraybackslash}m{0.22\linewidth}<{}
    }
         Array geometry & $\mathrm{PSL_{SIMO/MISO}}$ [dB] & $\mathrm{PSL_{MIMO}} $ [dB]\\
         \hline
         ULA & -8.78 & -17.57 \\
         \hline
         UCA & -7.90 & -15.80 \\
         \hline
         URA & -17.57 & -35.13 \\
         \hline
         UPCA & -13.26 & -26.52 \\
         
    \end{tabular}
\end{table}

\section{Conclusion}
This paper presents and quantifies the gains offered by MIMO configuration in the NF sensing across different aperture geometries. The best sensing resolution in terms of beamdepth is offered by UCA. Meanwhile, the planar arrays, such as URA and UPCA, offer better performance in terms of PSL at the expense of offered resolution. Independent of the antenna array geometry, MIMO provides an approximate $\sqrt{2}$ improvement in both NF sensing resolution and maximum NF range as compared to the SIMO/MISO scenario. Furthermore, due to squaring of the AF, MIMO processing achieves a quadratic reduction of PSL, equivalent to twofold reduction in dB.

\bibliographystyle{IEEEtran}
\bibliography{biblio.bib}

\end{document}